\documentclass{emulateapj}
\usepackage{graphicx}
\usepackage{times,color}

\usepackage{amsmath}

\shorttitle{On CR production efficiency at SNR shocks}
\shortauthors{J. SHIMODA ET AL.}

\begin{document}


\title{
On cosmic-ray production efficiency at supernova remnant shocks propagating into realistic diffuse interstellar medium
}


\author{
Jiro Shimoda\altaffilmark{1}, 
Tsuyoshi Inoue\altaffilmark{2}, 
Yutaka Ohira\altaffilmark{1}, 
Ryo~Yamazaki\altaffilmark{1},
Aya~Bamba\altaffilmark{1},
Jacco~Vink\altaffilmark{3}
}
\altaffiltext{1}{Department of Physics and Mathematics, Aoyama-Gakuin University, Sagamihara, Kanagawa 252-5258, Japan; s-jiro@phys.aoyama.ac.jp}
\altaffiltext{2}{National Astronomical Observatory of Japan, Mitaka, Tokyo 181-8588, Japan }
\altaffiltext{3}{Astronomical Institute Anton Pannekoek/Gravitation and AstroParticle Physics Amsterdam (GRAPPA),
University of Amsterdam, Science Park 904, 1098XH Amsterdam, The Netherlands}


\begin{abstract}
Using three-dimensional magnetohydrodynamics simulations, we show that the efficiency of cosmic-ray (CR) production in supernova remnants is over-predicted if it could be estimated based on proper motion measurements of H$\alpha$ filaments in combination with  shock-jump conditions.
Density fluctuations of the upstream medium cause shock waves to be rippled and oblique almost everywhere.
The kinetic energy of the shock wave is transferred into that of downstream turbulence as well as thermal energy related to the shock velocity component normal to the shock surface.
Our synthetic observation shows that the CR acceleration efficiency, as estimated from a lower downstream plasma temperature, is overestimated by 10-40\% because a rippled shock does not immediately dissipate all of the upstream kinetic energy. 
\end{abstract}


\keywords{
acceleration of particles
--- ISM: supernova remnants 
--- proper motions
--- shock waves
--- turbulence
}


\section{Introduction}
The energy density of Galactic cosmic rays (CRs) around the Earth is explained if 1--10~\% of supernova explosion energy is used for CR acceleration.
The CR production efficiency in the supernova remnants (SNRs) has been widely discussed and seems to be ubiquitously so high that the back reaction of CRs onto the background shock structure is significant.
One way to estimate the CR production efficiency is through a combination of measurements of the proper motion of the shock front and the temperature of shocked gas \citep[e.g.,][]{hughes00,tatischeff07,helder09,morlino13,morlino14}.
The expansion speed of the SNR has been measured in various wavelengths, from which the downstream temperature $T_{\rm proper}$ is predicted using the Rankine-Hugoniot shock jump condition.
If the actual downstream temperature $T_{\rm down}$ can be independently measured, 
then the CR production efficiency $\eta$ is given by
\begin{eqnarray}
\eta=\frac{T_{{\rm proper}}-T_{{\rm down}}}{T_{{\rm proper}}}~~,
\label{eta}
\end{eqnarray}
where we assume that all of the missing thermal energy goes into CR production.
Note that $\eta$ can be related to $\beta$ which was given by Equation~(22) of \citet{vink10},
as $\eta=1-\beta$.
Observations of the northeast region of the young SNR RCW 86 gives us an example.
The proper motion velocity of the synchrotron X-ray filaments is measured as $\sim6000\pm2800$~km~s$^{-1}$ \citep{helder09},  while those of the H$\alpha$ filaments range from 300 to 3000~km s$^{-1}$ with a mean of 1200~km~s$^{-1}$ \citep{helder13}.
Let the expansion speed of the SNR be 3000~km~s$^{-1}$ so that the X-ray and H$\alpha$ observations are consistent with each other.
If the proper motion velocity is equivalent to the shock velocity, then the downstream proton temperature is predicted by the Rankine-Hugoniot relation as $T_{\rm proper}= 17.6$~keV.
This value is different from the direct measurement, $T_{\rm down}=2.3\pm0.3$~keV, which is given by the line width of the broad component of the $\rm H\alpha$ emission \citep{helder09}.
Then, we obtain $\eta\approx 87$~\% which suggests extremely efficient CR acceleration.
Even if the shock velocity is as low as 1200~km~s$^{-1}$, the efficiency is 18~\%.
\par
In previous discussions, it was assumed that the shock was plane parallel --- that is, that the shock normal is parallel to the flow --- and that the measured proper motion velocity was identical to the shock velocity.
These assumptions would be suitable for a spherically symmetric shock wave propagating into a homogeneous medium.
However, they may not be true for actual SNRs.
The observed velocity of the proper motion of H$\alpha$ filaments is dispersed \citep[e.g.,][]{helder13},  which implies shock propagation through an inhomogeneous medium.
At present, it is widely accepted that the interstellar medium is highly inhomogeneous \citep[e.g.,][]{avillez07}, in particular, near the young SNRs \citep[e.g.,][]{fukui03,moriguchi05}.
So far, we have investigated the effects of upstream inhomogeneity and shown that various observational results can be explained \citep{inoue09,inoue10,inoue12,inoue13}.
Some predictions of magnetohydrodynamics (MHD) simulations regarding magnetic field amplification due to a turbulent dynamo downstream \citep{giacalone07,inoue09,inoue12,sano12}, have been observationally confirmed 
\citep{vink03,bamba03,bamba05a,bamba05b,yamazaki04,uchiyama07,sano13,sano14}.
In this paper, we will show that the above approximations may lead to overestimates of the CR production efficiency in SNRs.
In order to study the influence of upstream inhomogeneities, we perform a three-dimensional (3D) MHD simulation of a shock wave propagating into an inhomogeneous medium, and we simulate H$\alpha$ filaments whose proper motion is  synthetically measured.


\section{Shock Propagation Through An Inhomogeneous ISM}

\begin{figure}[htbp]
\begin{center}
\includegraphics[width=80mm]{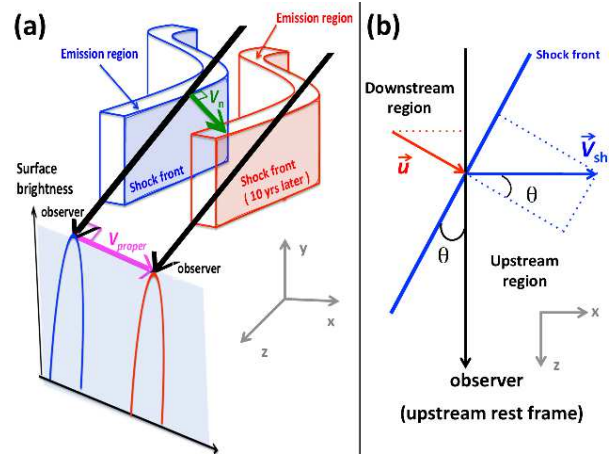}
\caption{
(a) Left panel  shows the relation between proper motion velocity and the velocity component normal to the shock surface for rippled shock.
The curved blue sheet represents a part of the rippled shock front and the emission region of H$\alpha$. 
The curved red sheet also represents those after a short-time propagation.
Thick black arrows show LOSs.
A Limb brightening effect causes a peaked profile in surface brightness as shown in the inner panel.
The magenta and green vectors represent the observed proper motion velocity $V_{{\rm proper}}$ and 
the velocity component normal to the shock surface $V_{{\rm n}}$, respectively.
One can see that $V_{{\rm proper}}$ is generally larger than $V_{{\rm n}}$.
(b) Right panel shows enlarged view of the local oblique shock.
The blue line represents the shock front that propagates along the $x$ axis, and the LOS direction (black arrow) is taken along the $z$ axis.
In the upstream rest frame, the downstream fluid velocity $\vec{u}$ is definitively parallel to the shock normal.
The $z$ component of $\vec{u}$ causes the Doppler shift in the broad H$\alpha$ line emission from this region.
}
\label{fig1}
\end{center}
\end{figure}

Multi-dimensional MHD simulations of shock propagation through an inhomogeneous diffuse ISM with a Kolmogorov-like density power spectrum have shown that the shock front is rippled due to the fluctuating inertia of the preshock ISM (see Giacalone \& Jokipii 2007 for the 2D case and Inoue et al. 2013 for the 3D case).
Their results strongly suggest that a SNR forward shock is locally oblique.
For oblique shocks, the downstream temperature is given by the velocity component normal to the shock surface $V_{\rm n}$ (not the shock velocity itself):
%
\begin{eqnarray}
k_{\rm B}T_{{\rm down}}=\frac{3}{16}m_{\rm p}V_{\rm n}^2~~,
\label{jump condition}
\end{eqnarray}
%
where $k_{\rm B}$ and $m_{\rm p}$ are the Boltzmann constant and proton mass, respectively.
The velocity measured by the proper motion is identical to the shock velocity component transverse to line of sight (LOS).
Thus, when the shock front is rippled, the proper motion velocity, $V_{\rm proper}$, can be larger than $V_{\rm n}$.
In Figure \ref{fig1}a, we illustrate this situation.
The blue curved sheet represents part of the rippled shock front emitting H$\alpha$ photons.
As seen in the bottom of the figure, the limb brightening effect causes a peaked profile in the surface brightness on the celestial sphere \citep{hester87}.
As the shock propagates, the peak of the  brightness moves outward (red sheet), which is observed as proper motion in the celestial sphere (magenta vector).
Since $V_{\rm proper}\geq V_{\rm n}$, the downstream temperature calculated based on the proper motion measurement can be overestimated, i.e., $T_{\rm proper}=3\,m_{\rm p} V_{\rm proper}^2/(16\,k_{\rm B})\geq T_{\rm down}$, so that $\eta$ is apparently non-zero in spite of no CR acceleration.
In the following, using the result of a 3D MHD simulation of a shock propagating through an inhomogeneous medium performed by \citet{inoue13}, we demonstrate that the above expectation is generally realized.


\section{Set up of MHD Simulation}

In this paper, we use the data from the simulation performed by \citet{inoue13}.
Here, we briefly summarize the set up of the simulation by \citet{inoue13}.
They studied shock propagation into an inhomogeneous medium that is parameterized by the amplitude 
of the density fluctuation $\Delta\rho/\left<\rho\right>_{0}$ assuming ideal MHD with adiabatic index $\gamma = 5/3$ and a mean molecular weight of 1.27, where $\left<\rho\right>_{0}$ is the initial mean density and $\Delta\rho\equiv(\left<\rho ^{2}\right>-\left<\rho\right>^{2}_{0})^{1/2}$ is the dispersion.
The fluctuations are given as a superposition of the sinusoidal functions with various wave numbers ($2\pi/L_{{\rm box}}\leq |k| \leq 256\pi/L_{\rm box}$).
The simulation is performed in a cubic numerical domain with volume $L_{{\rm box}}^3=(2{\rm\ pc})^{3}$ which is resolved by $(1024)^{3}$ unit cells.
The power spectrum of the density fluctuations is given by the isotropic power law, $P_{{\rm 1D}}(k)\equiv\rho_{k}^{2}k^{2}\propto k^{-5/3}$ for the above range of $k$, where $\rho_{k}$ is the Fourier component of the density.
The above Kolmogorov spectrum is consistent with the observed big-power-law-in-the-sky \citep{armstrong95}.
\par
The initial mean number density, thermal pressure, and magnetic field strength are set to be $\left<n\right>_{0}=0.5\ {\rm cm^{-3}}$, $P/k_{\rm B}=4\times 10^{3}\ {\rm K\ cm^{-3}}$, and $B_{0}=3.0\ {\rm \mu G}$, respectively.
These are the typical values in the diffuse ISM \citep{myers78,beck00}.
Thus, the initial mean sound speed and ${\rm Alfv\acute{e}n}$ velocity are $\left<c_{s}\right>=9.3\ {\rm km\ s^{-1}}$ and $\left<c_{A}\right>=8.2\ {\rm km\ s^{-1}}$, respectively.
To induce a blast wave shock, we set the hot plasma to $p_{\rm h}/k_{\rm B}=2\times10^{8}\ {\rm K\ cm^{-3}}$, $n_{\rm h}=0.05\ {\rm cm^{-3}}$, and $B_{\rm h}=3.0\ {\rm \mu G}$ at the $x=0$ boundary plane.
The resulting mean propagation speed of the shock is 1800 km s$^{-1}$, which is suitable for studying young SNRs, although the local shock velocity has a large dispersion due to the shock rippling.
\par
In this paper, we use data from the simulation for $\Delta\rho/\left<\rho\right>_{0}=0.3$, which can be regarded as a typical ISM model.
The reason for this is as follows. 
If we suppose that the turbulence in the ISM is driven by supernovae, then the driving scale of the turbulence and the degree of density fluctuation at the driving scale would be given as $L_{{\rm inj}}\sim100\ {\rm pc}$ and $\Delta\rho|_{L_{{\rm inj}}}/\left<\rho\right>_{0}\sim 1$, respectively \citep[e.g.,][]{avillez07}.
In that case, the degree of small-scale density fluctuations due to the cascade of the turbulence at the scale $L_{{\rm box}}=2\ {\rm pc}$ is estimated as $\Delta\rho|_{L_{{\rm inj}}}/\left<\rho \right>_{0}\simeq(L_{{\rm box}}/L_{{\rm inj}})^{1/3}\sim0.3$. 


\section{Results of MHD Simulation and Synthetic Observations}

\begin{figure}[htb]
\centering
\includegraphics[width=80mm]{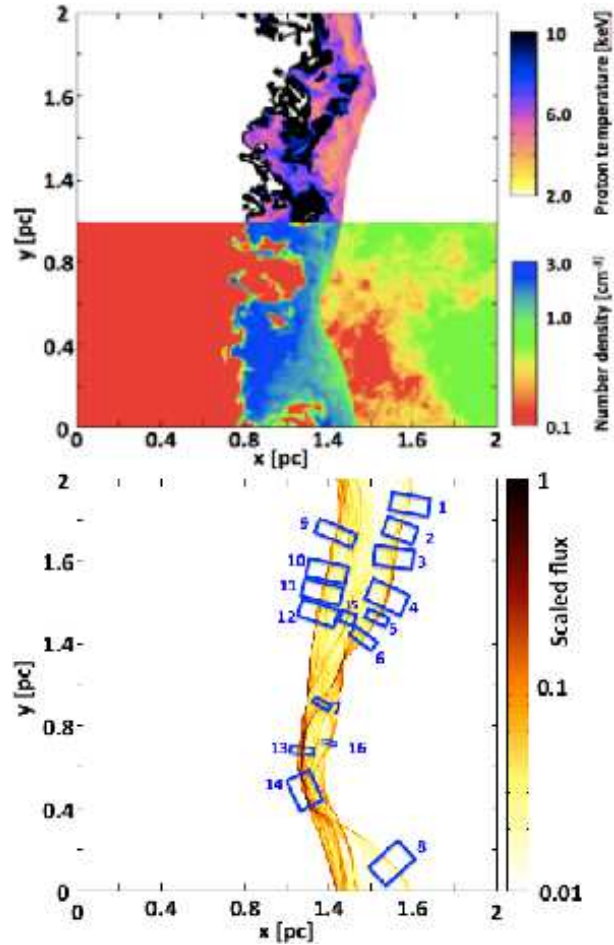}
\caption{
(Top panel)
two-dimensional slice of the proton temperature ({\it upper half}) and number density ({\it lower half}) in the $t_{{\rm obs}} = 700$ {\rm yr} and $z = 0$ \rm{pc} plane.
(Bottom panel) simulated H$\alpha$ image. 
We set the LOS along the z axis.
Color represents the scaled flux of H$\alpha$. 
We selected 16 local regions ({\it blue box}) in which the proper motion of the H$\alpha$ filament is measured to predict the downstream proton temperature.}
 \label{fig2}
\end{figure}

\begin{figure}[htb]
\centering
\includegraphics[width=80mm]{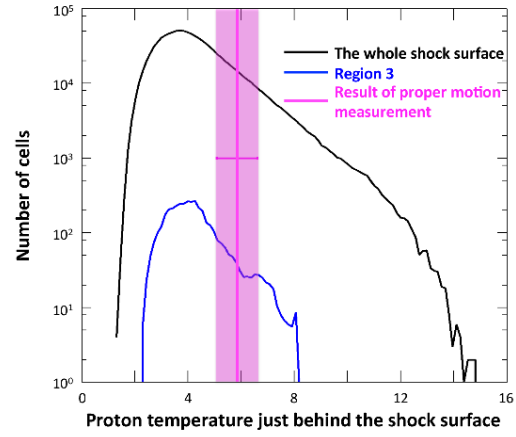}
\caption{
Distributions of the proton temperature of the fluid cell just behind the shock surface.
The black curve is for the whole shock surface, while the blue is for the surface of the LOS crossing the H$\alpha$ filament of Region~3.
The vertical magenta line represents the downstream proton temperature, $T_{\rm proper}$, which is inferred from the best-fit proper motion velocity with magenta belt showing the associated error.
}
\label{fig3}
\end{figure}

In this section, we show our simulation results.
The top panel of Figure \ref{fig2} represents the two-dimensional slice of the proton temperature ({\it upper half}) and the number density ({\it lower half}) in the $t_{{\rm obs}} = 700$ {\rm yr} and $z=0$ pc plane.
The proton temperature is estimated from $T_{\rm p}=P/(\rho k_{\rm B})$, where $P$ and $\rho$ are the pressure and density.
From Figure \ref{fig2}, we observe that shock waves propagate into the realistic ISM with various angles and velocities.
As a result, the temperature distribution is inhomogeneous (see the black curve of Figure \ref{fig3}).
In this inhomogeneous system, it may be that the relation between the proper motion velocity of H$\alpha$ and the downstream temperature does not satisfy the Rankine-Hugoniot relations.
\par
In order to estimate the deviation from the Rankine-Hugoniot relations, we calculate the proper motion of the H$\alpha$ emission over 10~yr (from $t_{{\rm obs}}=700$ to 710~yr). 
The H$\alpha$ emission observed from SNRs sometimes has narrow and broad components. 
The former is a characteristic of the cold interstellar medium which arises from direct excitation of the neutral hydrogen atoms crossing the shock surface.
The latter is a characteristic of the thermal broadening of the shocked protons which arises from hot hydrogen atoms generated by a charge exchange reaction between cold neutrals and the shocked protons.
In this paper, we calculate only the narrow component of the H$\alpha$ emission from our MHD simulation because the broad component of H$\alpha$ is not necessarily observed from SNRs. 
We consider hydrogen atoms, electrons, and protons as particles.
The results of our MHD simulation are valid as long as the ionization fraction of the upstream gas is sufficiently high because our MHD simulation does not take into account the ionization of neutral hydrogen atoms. 
Recent studies have shown that the ionization of neutral hydrogen atoms changes the collisionless shock structures \citep{ohira09,ohira10,ohira12,ohira13,ohira14,blasi12}. 
Therefore, we consider a highly ionized upstream gas in this paper because the effects of ionization can be neglected.
\par
According to \citet{heng07}, the rate (in units of ${\rm s^{-1}}$) that hydrogen atoms (denoted $\rm H$) will have a reaction $X$ ($X=E,~I,$ and CE for excitation, ionization, and charge exchange) with particles of type $s$ ($s=e$ and $p$ for electrons and protons, respectively) is given by
%
\begin{eqnarray}
R_{X,s}=n_{s}\int d^{3}\vec{v}_{\rm H}\int d^{3}\vec{v}_{s}f_{\rm H}(\vec{v}_{\rm H})
f(\vec{v}_{s})\Delta v_{s}\sigma _{X,s}(\Delta v_s)~~,
\label{reaction}
\end{eqnarray}
%
where $n_{s}$, $\vec{v}_{s}$, and $f_{s}$ are the number density, the velocity, and the distribution function for {\it particle} $s$, respectively. 
The velocity and distribution function of hydrogen atoms are described as
$\vec{v}_{\rm H}$ and $f_{\rm H}$, respectively.
The relative velocity between neutral hydrogen atoms and {\it particle} $s$ is denoted by $\Delta v_{s}=\left|\vec{v}_{\rm H}-\vec{v}_{s}\right|$.
We assume that distribution functions for each particle are represented by
%
\begin{eqnarray}
f_{\rm H}&=&\delta(\vec{v}_{\rm H})~~, \\
\label{hydrogen distribution}
f_{s}&=&\left(\frac{m_{s}}{2\pi k_{\rm B}T_{s}(\vec{r})}\right)^{3/2}
\exp\left(-\frac{m_{s}v^{2}_{s}(\vec{r})}{2k_{\rm B}T_{s}(\vec{r})}\right)~~, 
\label{plasma distribution}
\end{eqnarray}
where $v_{s}(\vec{r})=\left|\vec{v}_{s}-\vec{u}(\vec{r})\right|$, and $\vec{u}(\vec r)$ is the downstream fluid velocity.
We regard the distribution function of hydrogen atoms as a Dirac delta function because we only consider the narrow component.
In addition, we assume that the distribution functions of protons and electrons are Maxwellian with $T_{\rm p}=P/(\rho k_{\rm B})$ and $T_{\rm e}=0.01T_{\rm p}$, respectively citep{ohira07,ohira08,rakowski08}. 
In order to calculate equation (\ref{reaction}), we use our MHD simulation data for $u(\vec{r})$, $T_{\rm p}(\vec{r})$, and $n_{s}(\vec{r})$, as well as the data of \citet{janev93} for cross sections.
Calculating the excitation rate, $R_{\rm E}=R_{\rm E,p}+R_{\rm E,e}$, for each cell of our MHD simulation and integrating $n_{\rm H}(\vec r)R_{\rm E}(\vec r)$ along the LOS (z axis), we obtain the surface emissivity of the H$\alpha$ emission, 
%
\begin{eqnarray}
S(x,y)=\int n_{\rm H}(\vec r)R_{\rm E}(\vec r)dz~~,
\label{surface brightness}
\end{eqnarray}
%
where we consider only direct excitation from the ground state to the $n=3$ level.
Neutral hydrogen atoms are ionized in the downstream region. 
The density of neutral hydrogen atoms in the downstream region is given by
%
\begin{eqnarray}
&&
 n_{\rm H}(\vec{r})=n_{{\rm H,0}}(\vec{r})
 \exp\left[-R_{\rm I}(\vec{r})(t_{\rm{obs}}-t_{{\rm sh}}(\vec{r}))\right]~~, 
 \label{ionization} 
\end{eqnarray}
%
where $R_{\rm I}(\vec{r})=R_{\rm I,p}(\vec{r})+R_{\rm CE,p}(\vec{r})+R_{\rm I,e}(\vec{r})$ and $t_{\rm sh}(\vec{r})$ is the time when the shock wave passes though point $\vec{r}$.
For the initial hydrogen density, $n_{\rm H}$, we assume that the initial ionization fraction of the ISM is uniform.
The bottom panel of Figure \ref{fig2} is the H$\alpha$ image obtained from equations~(\ref{reaction})--(\ref{ionization}).
\par
As shown in the bottom  panel of Figure \ref{fig2}, we select 16 regions which contain clear filamentary structure of H$\alpha$ to measure proper motion.
We extract a surface brightness profile from these regions and analyze their proper motions in the same way as \citet{helder13}.
To measure the proper motion, we shift the normalized profiles over one another in steps of one bins, calculating the $\chi^2$ values for each shift.
The length of one bin is taken as $1.9\times10^{15}$cm, which is comparable to the angular resolution of the optical instrument for the typical distance to the source of a few kiloparsecs.
The best-fitting shift is determined by fitting a parabola to the three $\chi^2$ values surrounding the minimal $\chi^2$.
We estimate the 1$\sigma$ uncertainties on the best-fit proper motion, which correspond to $\Delta\chi^2=1$.
Then, the best-fit proper motion velocity $V_{{\rm proper}}$ is related to $T_{{\rm proper}}$ as $T_{{\rm proper}}=3m_{\rm p} V_{\rm proper}^2/16k_{\rm B}$, which is the same way as in previous actual observational studies.
\par
In order to evaluate $\eta$ from equation~(\ref{eta}), we calculate the downstream proton temperature $T_{{\rm down}}$ in two ways.
First, we take $T_{{\rm down}}$ as the mean of the downstream temperatures of fluid cells just behind the shock surface of the LOS crossing the H$\alpha$ filament (Case~1).
As a typical example, in Figure~\ref{fig3}, we show the distributions of the proton temperature $T_{\rm down}$ of the fluid cell just behind the shock surface; the black curve represents for the whole shock surface, while the blue represents for the surface on the LOS crossing the H$\alpha$ filament of Region~3.
Note that the proper motion velocity of Region~3 corresponds to the mean of our 16 regions (see Table~\ref{tb1}). 
The vertical magenta line represents  $T_{\rm proper}$ for Region~3 with the magenta belt showing the associated error.
The value of $T_{\rm proper}$ is higher than the mean of $T_{\rm down}$.
We calculate, from equation~(\ref{eta}), the apparent CR production efficiency as $\eta\sim0.3\pm0.1$.
\par
Next, we consider a situation similar to actual observations where the downstream temperature $T_{\rm down}$ is estimated from the line width of the broad H$\alpha$ component (Case~2).
It is hard to calculate exactly the broad emission component in our model. 
Instead, we perform a simple, approximate calculation.
We obtain $T_{{\rm down}}$ from the FWHM of the sum of the shifted Maxwellian weighted by the brightness of the broad H$\alpha$ component, $n_{{\rm H,0}}(\vec{r})\xi_{{\rm CE}}(\vec{r})$, of the fluid cell at $\vec{r}$ just behind the shock surface on the LOS crossing the H$\alpha$ filament, where $\xi_{{\rm CE}}(\vec{r})=R_{{\rm CE,p}}(\vec{r}) [R_{{\rm I,p}}(\vec{r})+R_{{\rm I,e}}(\vec{r})]^{-1}$.
Then, we find that $\eta$ becomes slightly larger than for Case~1 (Table \ref{tb1}).
This is because the hot hydrogen atom emitting the broad H$\alpha$ component is generated by a charge exchange reaction, whose  cross-section decreases rapidly if the relative velocity is higher than $\approx2000$~km~s$^{-1}$. 
Thus, the observed downstream temperature may be biased against the particular temperature.
This effect has already been investigated for a one-dimensional shock wave through detailed analysis of H$\alpha$ emission \citep{vanadelsberg08}.
Since a rippled shock front generates dispersion in the downstream fluid velocity, one might consider that the line width to be spread by  downstream bulk motion.
If this Doppler effect were significant, then the measured $T_{\rm down}$ would tend to be higher than the actual downstream proton temperature, resulting in lower $\eta$.
However, this is not the case for our present synthetic observation.
It is known for oblique shocks that in the upstream rest frame, 
the downstream fluid velocity $\vec{u}$ is parallel to the shock normal 
(Figure~\ref{fig1}b).
Hence, if the rippled shock front is viewed nearly edge-on as in the present case, then the Doppler broadening is not so significant.
\par
Table~\ref{tb1} shows the measured proper motion velocity and apparent CR production 
efficiency $\eta$ for the 16 regions.  
As expected in section~2, $T_{{\rm proper}}$ is higher than $T_{\rm down}$ and the efficiency $\eta$ is positive, even though our simulations do not involve the effects of comic-ray acceleration.



\section{Discussion}
We have shown that the CR production efficiency $\eta$ seems to be overestimated in the shock wave of SNRs propagating into a realistic ISM if the post-shock temperature $T_{\rm proper}$ is estimated from the proper motion of the H$\alpha$ filaments in combination with the Rankine-Hugoniot relation for a plane-parallel shock.
It may not be a suitable assumption for actual SNR shocks that the shock wave is plane parallel and that the measured proper motion velocity is equivalent to the shock velocity.
Density fluctuations of a realistic ISM make the rippled, locally oblique shock front almost everywhere.
For the oblique shocks, the post-shock temperature is given not by the shock velocity itself but by the velocity component normal to the shock surface $V_{\rm n}$ as shown by equation~(\ref{jump condition}).
Because proper motion measurements give us the velocity component transverse to the LOS (see Figure~\ref{fig1}), the predicted post-shock temperature $T_{{\rm proper}}$ given by the Rankine-Hugoniot relation with the assumption of a plane parallel shock is larger than actual downstream temperature $T_{{\rm down}}$.
Therefore, we claim that the CR production efficiency $\eta$ has some uncertainty and can be positive (up to 0.4 in our case) despite no CR acceleration.
\par
As shown in the Appendix,  a simple analytical argument gives the upper and lower bounds of $\eta$ as
%
\begin{equation}
\left(\frac{\Delta\rho}{\left<\rho\right>_{0}}\right)^2
\lesssim\eta\lesssim 2\frac{\Delta\rho}{\left<\rho\right>_{0}}~~,
\end{equation}
where  $\Delta\rho/\left<\rho\right>_{0}$ is the upstream density fluctuation at the scale $L_{\rm box}=2$~pc.
Since we have set $\Delta\rho/\left<\rho\right>_{0}=0.3$ in our present simulation study, this analytical formula is roughly consistent with our numerical result.
RCW~86 is likely a SNR expanding in the windblown bubble \citep{vink97,vink06}, and a part of the shock collided with dense clumps and/or a cavity wall very recently \citep{yamaguchi08} so that we expect a larger value of $\Delta\rho/\left<\rho\right>_{0}$ than that of the ISM.
If the CR acceleration is inefficient so that the nonlinear effect can be neglected, then we expect $\Delta\rho/\left<\rho\right>_{0}\approx0.4$ in order to explain the observational result $\eta=0.2$--0.9 (see section~1).
On the other hand, shock deformation in $\sim10$~pc scale may be smaller for SNRs such as SN~1006, Tycho's remnant, and SNR~0509--67.5, which are embedded in the ISM with smaller $\Delta\rho/\left<\rho\right>_{0}$ than RCW~86.
The global  H$\alpha$ image of SN~1006, whose radius is about 10~pc, looks like a circular ring except for the northwest region, while smaller-scale ($\lesssim$ a few parsecs) rippling can also be seen \citep{raymond07,winkler14}.
Tycho's remnant has a radius of about 3~pc, and its whole H$\alpha$ shape is no longer circular \citep{raymond10}.
Indeed, several observational results have indicated the inhomogeneity of the ambient medium around SN~1006 \citep{dubner02,raymond07,miceli14} and Tycho's remnant \citep{reynoso99,Ishihara10}.
SNR~0509--67.5 is also round in shape with a radius of 3.6~pc, however, the southwest part of the remnant is rippled and has many H$\alpha$ filaments \citep[see, e.g., Fig.~1 of][]{helder10}.
These observational results for H$\alpha$ morphology on a few parsecs or smaller scale are consistent with our model with a typical ISM density fluctuation.
Therefore, we should still pay attention to the effect of upstream inhomogeneity when the CR acceleration efficiency is discussed in these remnants.
In order to reproduce the morphology of the whole remnant and the smaller-scale structure simultaneously, we require larger-scale simulation, keeping the same spatial resolution as in the present study, which is currently difficult due to the limitation of computer resources and must remains a future work.

\par
At oblique shocks, the upstream velocity component parallel to the shock front is not dissipated across the shock. 
For the case of an edge-on view of the rippled shock, such a component mainly turns out to be transverse to the LOS in the downstream region (see Figure~1(a)), so that it becomes an unseen, missing component--- it does not even contribute to the width of the broad H$\alpha$ line.
In previous observational arguments, the missing energy was attributed to CR acceleration. 
In the present case, the post-shock fluid stream lines become "turbulence" after the crossing time of the shock rippling scale (the driving scale of "turbulence"). 
Note that this driving scale $\sim$ 0.1 pc (Inoue et al. 2013) is much larger than the typical width of the emission region, indicating that turbulent line broadening cannot be measured by H$\alpha$ emission.
Since the downstream turbulence is created by the effect of the rippled shock wave \citep{giacalone07}, the induction of the turbulence can be understood as a consequence of Crocco's theorem in hydrodynamics.  
The strength of the induced turbulence depends on the degree of the density inhomogeneity in the pre-shock medium.  
In \citet{inoue13}, we found that the velocity dispersion of the turbulence can be well described by a formula obtained from the modified growth velocity of the Richtmyer--Meshkov instability as a function of the upstream density dispersion.
%
\par
One can also find from the synthetic H$\alpha$ image (bottom panel of Figure~\ref{fig2}) that regions~1--6  precede regions~9--12, so that one might think that the proper motion velocities of regions~1--6 are higher than those of regions~9--12.
However,  this is not true (see Table~1). 
We find the outermost parts of the H$\alpha$ filament do not always have the fastest shock velocity or the highest downstream temperature.
This is because the shock front has effective surface tension and is stable with respect to the rippling deformations. 
Thus, even though some regions of the shock front are decelerated (accelerated) due to the passage of the dense (thin) region, they will be accelerated (decelerated) once the dense (thin) region passes into 
the downstream region. 
\par
In the present analysis, we have seen $\eta\geq0$ (that is, $T_{{\rm proper}}> T_{{\rm down}}$) for all 16~regions, which implies that the proper motion velocity $V_{\rm proper}$ is larger than the velocity component normal to the shock surface $V_{{\rm n}}$.
We have set our LOS orthogonal to the global direction of the shock propagation.
However, when the shock wave propagates nearly toward us (along the LOS), $V_{\rm proper}$ can be smaller than $V_{{\rm n}}$.
For example, \citet{salvesen09} measured the proper motion velocity of H$\alpha$ filaments of the Cygnus Loop and simultaneously derived the downstream gas temperature from the thermal X-ray spectrum there.
Then, they obtained the fraction of the CR pressure $P_{{\rm CR}}$ to the thermal gas pressure $P_{\rm G}$ in the downstream region.
According to their analysis, many H$\alpha$ filaments have $P_{{\rm CR}}/P_{\rm G}\leq0$.
Since they assumed  a strong shock with a compression ratio of 4, an adiabatic index of 5/3, and a temperature equilibrium of $T_{e}=T_{i}=T_{\rm down}$, the ratio $P_{{\rm CR}}/P_{\rm G}$ is related to $\eta$ as
\begin{eqnarray}
\frac{P_{{\rm CR}}}{P_{\rm G}}=\frac{\eta}{1-\eta}~~.
\end{eqnarray}
Here $T_{e}$ and $T_{i}$ are downstream electron and ion temperatures, respectively. 
Since $\eta>1$ is unphysical, $P_{{\rm CR}}/P_{\rm G}<0$ means $\eta<0$, that is, $T_{{\rm proper}}< T_{{\rm down}}$.
Therefore, the observational result for the Cygnus Loop might be explained by our model.
Moreover, it is suggested that the proper motion velocity is underestimated due to the shock obliqueness.


\acknowledgments
We thank the referee for useful comments and suggestions.
Numerical computations were carried out on the XC30 system at the Center for Computational Astrophysics (CfCA) of National Astronomical Observatory of Japan and K computer at the RIKEN Advanced Institute for Computational Science (No. hp120087).
This work is supported by Grant-in-aids from the Ministry of Education, Culture, Sports, Science, and Technology (MEXT) of Japan, No. 23740154 (T.I.) and No. 248344 (Y.O.), No. 22684012 (A.B.).
T.I. and R.Y. deeply appreciate Research Institute, Aoyama-Gakuin University for helping our research by the fund.
R.Y. also thanks ISSI (Bern) for support of the team ``Physics of the Injection of Particle Acceleration at 
Astrophysical, Heliospheric, and Laboratory Collisionless Shocks.'' 

\appendix

\section{Analytical Estimate of $\eta$}

We present a simple analytical argument which relates $\eta$ to the upstream density fluctuation.
We simplify the upstream medium as a mixture of two components: overdense clumps with density $\left<\rho\right>_{0}+\Delta\rho$ and underdense gas with density $\left<\rho\right>_{0}-\Delta\rho$.
The characteristic size of the clumps $\lambda$ is the same as their separation.
In the present case, $\lambda$ is on the order of $L_{\rm box}=2$~pc.
Initially, the planar shock surface collides with the clumps.
Its propagation speed in the clumps $V_+$ is slower
than that in the underdense gas $V_-$, so that the shock surface is deformed.
Assuming momentum conservation, these are related as 
%
\begin{equation}
(\left<\rho\right>_{0}+\Delta\rho)V_+^2
\approx (\left<\rho\right>_{0}-\Delta\rho)V_-^2~~.
\end{equation}
%
Then, we find  $V_+/V_-\approx 1-\Delta\rho/\left<\rho\right>_{0}$ for small $\Delta\rho/\left<\rho\right>_{0}$.
As the shock front goes ahead a distance $\lambda$ in the underdense gas, the shock surface in the clumps is left at a distance $\delta=\lambda - (\lambda/V_-)V_+$ behind the preceding surface in the underdense gas.
Therefore, one derives the deformation angle $\theta$, which is an angle between the shock velocity $\vec{V}_{\rm sh}$ and the shock normal (see the right panel of Figure~1), as $\theta\approx\delta/\lambda\approx\Delta\rho/\left<\rho\right>_{0}$.
\par
The downstream temperature $T_{\rm down}$ is predominantly determined by  the overdense clump, that is, Equation~(\ref{jump condition}) with $V_{\rm n}=V_+\cos\theta$.
For our present geometry, in which shock surfaces are viewed nearly edge-on, the proper motion velocity $V_{\rm proper}$ is roughly equal to the shock velocity $V_{\rm sh}$.
If we observe the proper motion velocity of the shock surface propagating into the overdense clump, then 
$V_{\rm proper}\approx V_+$, while $V_{\rm proper}\approx V_-$ for the shock propagation into the
underdense gas.
Hence, we find $\eta=1-\cos^2\theta\approx (\Delta\rho/\left<\rho\right>_{0})^2$ for the former case and $\eta=1-(V_+/V_-)^2\cos^2\theta\approx2\Delta\rho/\left<\rho\right>_{0}$ for the latter case.
In a more complicated case of our present simulation study, we expect $V_+\lesssim V_{\rm sh}\lesssim V_-$ and these values of $\eta$ may give lower and upper bounds, so that $(\Delta\rho/\left<\rho\right>_{0})^2\lesssim\eta\lesssim 2\Delta\rho/\left<\rho\right>_{0}$. 


%
\begin{deluxetable}{c|cc|cc|cc}
\tablecaption{calculation results for selected 16 regions}
\tablehead{
& & & \multicolumn{2}{|c|}{Case~1} & \multicolumn{2}{c}{Case~2} \\ 
Region           & $V_{{\rm proper}}$ & $T_{{\rm proper}}$ &
$T_{{\rm down}}$ & $\eta$             & $T_{{\rm down}}$   & $\eta$ \\
& ($10^{8}$cm~s$^{-1}$) & (keV) & (keV) & & (keV) & 
}
\startdata
1 & 1.8$\pm$0.1 & 6.0$\pm$0.5 & 5.7 & 0.06$\pm$0.07 & 5.5 & 0.08$\pm$0.07\\
2 & 1.8$\pm$0.0 & 6.6$\pm$0.3 & 6.0 & 0.09$\pm$0.04 & 5.7 & 0.14$\pm$0.04\\
3 & 1.7$\pm$0.1 & 5.8$\pm$0.8 & 4.2 & 0.27$\pm$0.09 & 3.9 & 0.33$\pm$0.08\\
4 & 1.4$\pm$0.1 & 3.7$\pm$0.4 & 3.0 & 0.19$\pm$0.08 & 2.9 & 0.22$\pm$0.07\\
5 & 1.4$\pm$0.1 & 3.7$\pm$0.4 & 2.6 & 0.28$\pm$0.08 & 2.6 & 0.29$\pm$0.08\\
6 & 1.4$\pm$0.0 & 4.0$\pm$0.2 & 2.9 & 0.28$\pm$0.04 & 2.8 &  0.30$\pm$0.03\\
7 & 1.4$\pm$0.1 & 3.9$\pm$0.6 & 3.3 & 0.20$\pm$0.10 & 3.1 & 0.20$\pm$0.11\\
8 & 1.7$\pm$0.1 & 5.4$\pm$0.6 & 4.1 & 0.24$\pm$0.08 & 3.9 &  0.28$\pm$0.08\\
9 & 2.0$\pm$0.1 & 7.5$\pm$0.9 & 5.1 & 0.32$\pm$0.07 & 4.9 & 0.34$\pm$0.07\\
10 & 2.1$\pm$0.0 & 8.2$\pm$0.2 & 5.5 &0.33$\pm$0.02 & 5.2 &  0.36$\pm$0.02\\
11 & 2.0$\pm$0.0 & 8.2$\pm$0.3 & 5.2 & 0.37$\pm$0.02 & 5.0 & 0.39$\pm$0.02\\
12 & 1.9$\pm$0.1 & 7.2$\pm$0.6 & 4.3 & 0.40$\pm$0.05 & 4.3 & 0.41$\pm$0.05\\
13 & 1.6$\pm$0.1 & 5.2$\pm$0.5 & 3.5 & 0.33$\pm$0.06 & 3.4 & 0.35$\pm$0.06\\\
14 & 1.6$\pm$0.1 & 5.0$\pm$0.6 & 4.3 & 0.10$\pm$0.10 & 3.9 & 0.22$\pm$0.09\\
15 & 1.4$\pm$0.1 & 4.0$\pm$0.4 & 2.9 & 0.27$\pm$0.07 & 2.9 & 0.29$\pm$0.07\\
16 & 2.0$\pm$0.0 & 8.2$\pm$0.1 & 7.2 & 0.11$\pm$0.01 & 7.7 & 0.06$\pm$0.01\\ \hline
Mean/std. dev. & 1.7/0.24 & 5.8/1.6 & 4.4/1.3 & 0.24/0.10 & 4.2/1.3 & 0.27/0.10 \\
\enddata
\label{tb1}
\end{deluxetable}
%

%
%

%
%

%
%


\begin{thebibliography}{}
\bibitem[Armstrong et al.(1995)]{armstrong95} 
Armstrong, J. W., Rickett, B. J., \& Spangler, S. R. 1995, \apj, 443, 209

\bibitem[Bamba et al.(2003)]{bamba03} 
Bamba, A., et al. 2003, ApJ, 589, 827

\bibitem[Bamba et al.(2005a)]{bamba05a} 
Bamba, A., et al. 2005a, ApJ, 621, 793

\bibitem[Bamba et al.(2005b)]{bamba05b} 
Bamba, A., Yamazaki, R., \& Hiraga, J. S. 2005b, ApJ, 632, 294

\bibitem[Beck(2000)]{beck00}
Beck, R. 2000, Space Sci. Rev., 99, 243

\bibitem[Blasi et al.(2012)]{blasi12}
Blasi, P., Morlino, G., Bandiera R., Amato, E., \& Caprioli, D., 2012, \apj, 755, 121 
 
\bibitem[de Avillez \& Breitschwerdt(2007)]{avillez07}
de Avillez, M. A., \& Breitschwerdt, D. 2007, \apj, 665, 35

\bibitem[Dubner et al.(2002)]{dubner02}
Dubner, G. M., Giacani, E. B., Goss W. M., et al. 2002, A\&A, 387, 1047


\bibitem[Fukui et al.(2003)]{fukui03}
Fukui, Y. et al. 2003, PASJ, 55, L61

\bibitem[Giacalone \& Jokipii(2007)]{giacalone07}
Giacalone, J., \& Jokipii, J. R. 2007, \apj, 663, 41

\bibitem[Helder et al.(2013)]{helder13} 
Helder, E. A., Vink, J., Bamba, et al. 2013, \mnras, 435, 910

\bibitem[Helder et al.(2010)]{helder10}
Helder, E. A., Kosenko, D. \& Vink, J. 2010, ApJ, 719, L140

\bibitem[Helder et al.(2009)]{helder09} 
Helder, E.A., Vink, J., Bassa, C. G. et al. 2009, Sci, 325, 719

\bibitem[Heng \& McCray(2007)]{heng07}
Heng, K., \& McCray, R. 2007, \apj, 654, 923

\bibitem[Hester(1987)]{hester87}
Hester, J. J. 1987, ApJ, 314, 187

\bibitem[Hughes et al.(2000)]{hughes00}
Hughes, J. P., Rakowski, C. E., \& Decourchelle, A. 2000, ApJ, 543, L61

\bibitem[Inoue et al.(2009)]{inoue09}
Inoue, T., Yamazaki, R., \& Inutsuka, S. 2009, ApJ, 695, 825

\bibitem[Inoue et al.(2010)]{inoue10}
Inoue, T., Yamazaki, R., \& Inutsuka, S. 2010, ApJ, 723, L108

\bibitem[Inoue et al.(2012)]{inoue12}
Inoue, T., Yamazaki, R., Inutsuka, S., \& Fukui, Y. 2012, ApJ, 744, 71

\bibitem[Inoue et al.(2013)]{inoue13}
Inoue, T., Shimoda, J., Ohira, Y., \& Yamazaki, R. 2013, \apj, 772, L20

\bibitem[Ishihara et al.(2010)]{Ishihara10}
Ishihara, D., Kaneda, H., Fukuzawa, A. et al. 2010 A\&A, 521, L61

\bibitem[Janev \& Smith(1993)]{janev93}
Janev, R. K., \& Smith, J. J. 1993, Cross Sections for Collision Processes of Hydrogen Atoms with Electrons, Protons and Multiply Charged Ions (Vienna: Int. At. Energy Agency)

\bibitem[Miceli et al.(2014)]{miceli14}
Miceli, M., Acero, F., Dubner, G. et al. 2014, ApJ, 782, L33

\bibitem[Morlino et al.(2014)]{morlino14}
Morlino, G., Blasi, P., Bandiera, R., \& Amato, E. 2014 A\&A 562, A141

\bibitem[Morlino et al.(2013)]{morlino13}
Morlino, G., Blasi, P., Bandiera, R., \& Amato, E. 2013 A\&A 557, A142

\bibitem[Moriguchi et al.(2005)]{moriguchi05}
Moriguchi, Y. et al. 2005, ApJ, 631, 947

\bibitem[Myers(1978)]{myers78} 
Myers, P. C. 1978, \apj, 225, 380

\bibitem[Ohira \& Takahara(2007)]{ohira07} 
Ohira, Y., \& Takahara, F., 2007, \apj, 661, L171

\bibitem[Ohira \& Takahara(2008)]{ohira08} 
Ohira, Y., \& Takahara, F., 2008, \apj, 688, 320

\bibitem[Ohira et al.(2009)]{ohira09} 
Ohira, Y., Terasawa, T., \& Takahara, F., 2009, \apj, 703, L59

\bibitem[Ohira \& Takahara(2010)]{ohira10} 
Ohira, Y., \& Takahara, F., 2010, \apj, 721, L43

\bibitem[Ohira(2012)]{ohira12} 
Ohira, Y., 2012, \apj, 758, 97

\bibitem[Ohira(2013)]{ohira13} 
Ohira, Y., 2013, \prl, 111, 245002

\bibitem[Ohira(2014)]{ohira14} 
Ohira, Y., 2014, \mnras, 440, 514

\bibitem[Rakowski et al.(2008)]{rakowski08} 
Rakowski, C. E., Laming, J. M., \& Ghavamian, P., 2008, \apj, 684, 348

\bibitem[Raymond et al.(2007)]{raymond07}
Raymond, J.C., Korreck, K.E., Sedlacek, Q.C., Blair, W. P. et al., 2007 ApJ, 659, 1257

\bibitem[Raymond et al.(2010)]{raymond10}
Raymond, J.C., Winkler, P. F., Blair, W. P. et al., 2010 ApJ, 712, 901

\bibitem[Reynoso et al.(1991)]{reynoso99}
Reynoso, E. M., Vel$\acute{a}$zquez, P. F., Dubner, G. M., \& Goss, W. M. 1991, AJ, 117, 1827

\bibitem[Sano et al.(2012)]{sano12}
Sano, T., Nishihara, K., Matsuoka, C. \& Inoue, T. 2012, ApJ, 758, 126

\bibitem[Sano et al.(2013)]{sano13}
Sano, H., Tanaka, T., Torii, K., Fukuda, T. et al. 2013, ApJ, 778, L59 

\bibitem[Sano et al.(2014)]{sano14}
Sano, H., Fukuda, T., Yoshiike, S., Sato, J. et al. 2014, arXiv:1401.7418

\bibitem[Salvesen et al.(2009)]{salvesen09}
Salvesen, G., Raymond, J., C., \& Edgar, R., E. 2009 \apj, 702, 327

\bibitem[Tatischeff \& Hernanz(2007)]{tatischeff07}
Tatischeff, V. \& Hernanz, M. 2007, ApJ, 663, L101

\bibitem[Uchiyama et al.(2007)]{uchiyama07}
Uchiyama, Y. et al. 2007, Nature, 449, 576

\bibitem[van Adelsberg et al.(2008)]{vanadelsberg08}
van Adelsberg, M. et al. 2008, ApJ, 689, 1089

\bibitem[Vink et al.(1997)]{vink97}
Vink, J. et al. 1997, A\&A, 328, 628 

\bibitem[Vink et al.(2006)]{vink06}
Vink, J. et al. 2006, ApJ, 648, L33 

\bibitem[Vink \& Laming(2003)]{vink03}
Vink, J., \& Laming, J. M. 2003, ApJ, 584, 758

\bibitem[Vink et al.(2010)]{vink10}
Vink, J., Yamazaki, R., Helder, E.A. et al. 2010, ApJ, 722, 1727

\bibitem[Winkler et al.(2014)]{winkler14}
Winkler, P. F. et al. 2014, ApJ, 781, 65

\bibitem[Yamaguchi et al.(2008)]{yamaguchi08}
Yamaguchi, H., Koyama, K., Nakajima, H. et al. 2008, PASJ, 60, S123

\bibitem[Yamazaki et al.(2004)]{yamazaki04}
Yamazaki, R., Yoshida, T., Terasawa, T. et al. 2004, A\&A, 416, 595

\end{thebibliography}
\end{document}